\documentclass[
twocolumn,amsmath]{aastex701}

\usepackage{color}
\usepackage[T1]{fontenc} 
\usepackage[utf8]{inputenc} 
\usepackage{amssymb}
\usepackage{float}
\usepackage{placeins}
\usepackage{hyperref}
\usepackage{natbib}
\usepackage{lipsum}
\usepackage{booktabs}
\usepackage{float}
\usepackage{subcaption}

\setcitestyle{notesep={ }}
\newcommand{\m}{\mathrm}
\begin{document}
	\title{Estimating Cosmological Parameters from Localized Fast Radio Bursts:  A Method for Removing Milky Way Dispersion-Measure Contributions} 
	
	\correspondingauthor{Hongwei Yu}
	\email{hwyu@hunnu.edu.cn}
	\correspondingauthor{Puxun Wu}
	\email{pxwu@hunnu.edu.cn}
	\author[0009-0009-3644-2082]{Yuchen Zhang}
	\affiliation{Department of Physics,  Key Laboratory of Low Dimensional Quantum Structures
		and Quantum Control of Ministry of Education, and Hunan Research Center of the Basic Discipline for Quantum Effects and Quantum Technologies, Hunan Normal University, Changsha, Hunan 410081, China}
	\email{}
	\author[0000-0003-2721-2559]{Yang Liu}
	\affiliation{Purple Mountain Observatory, Chinese Academy of Sciences, No. 10 Yuanhua Road, Nanjing 210023, China}
	\email{}
	\author[0000-0002-3303-9724]{Hongwei Yu} \email{hwyu@hunnu.edu.cn}
	\affiliation{Department of Physics,  Key Laboratory of Low Dimensional Quantum Structures
		and Quantum Control of Ministry of Education, and Hunan Research Center of the Basic Discipline for Quantum Effects and Quantum Technologies, Hunan Normal University, Changsha, Hunan 410081, China}

	\author[0000-0002-9188-7393]{Puxun Wu}
	\affiliation{Department of Physics,  Key Laboratory of Low Dimensional Quantum Structures
		and Quantum Control of Ministry of Education, and Hunan Research Center of the Basic Discipline for Quantum Effects and Quantum Technologies, Hunan Normal University, Changsha, Hunan 410081, China}
	\email{pxwu@hunnu.edu.cn}
	
\begin{abstract}

Fast radio bursts (FRBs) are emerging as  powerful probes for cosmology.  However, cosmological inference based on FRB dispersion measures (DMs) is limited by uncertainties in the Milky Way contribution, including those from the Galactic interstellar medium and the Galactic halo. In this Letter, we propose a method that  eliminates the Milky Way contribution by using DM differences between localized FRBs within the same sky region.  The method removes the need to adopt a specific Galactic electron-density model or a prior assumption for the Galactic halo DM. We validate the reliability of the method using mock FRB samples and show that it successfully recovers the fiducial cosmological parameter. Applying the method to current localized FRB data, we obtain a constraint on $\Gamma \equiv \Omega_b H_0 f_{\rm d}$ that differs from that inferred using the conventional treatment of the Milky Way contribution. This difference highlights the importance of Milky Way DM systematics in FRB cosmology and demonstrates the potential of differential DM methods for future large samples of localized FRBs.
  
\end{abstract}
	
\section{Introduction} 
Fast radio bursts (FRBs) are millisecond-duration radio transients,   typically observed at radiation frequencies of  order $\sim$GHz, first discovered by Lorimer et al. in archival pulsar survey data~\citep{Lorimer:2007qn}. As FRB signals travel from their sources to the observer,  they interact with free electrons long the line of sight,  producing a frequency-dependent arrival-time delay.  The corresponding dispersion measure (DM) accumulates with propagation distance and therefore encodes information about the distribution of ionized matter in the Universe. Localized FRBs with identified host galaxies and measured redshifts can thus be used as cosmological probes through their DM-redshift relation.

They have been used to explore the cosmic reionization history~\citep{Zheng:2014rpa, Caleb:2019apf, Beniamini:2020ane, Hashimoto:2021tyk, Wei:2024tpv, Shaw:2024slf}, constrain the photon rest mass~\citep{Wu:2016brq, Bonetti:2016cpo, Bonetti:2017pym, Shao:2017tuu, Xing:2019geq, Wei:2020wtf, Wang:2021nrl, Chang:2022qct, Lin:2023jaq, Wang:2023fnn, Lemos:2025qyh, Chang:2024hnn, Ran:2024avn, Wang:2024rgu, Zhang:2025rvu}, 
and determine cosmological parameters,  such as the Hubble constant $H_{0}$~\citep{Li:2017mek, hagstotz2022new, james2022measurement, Liu:2022bmn, Fortunato:2024hfm, Piratova-Moreno:2025cpc, Wu:2021jyk, Liu:2026epv, Zhuge:2025urk, Sales:2025shu, Jia:2025kvp, Gao:2025fcr, Wang:2025ugc, Xu:2025ddk, Lemos:2026snc}, the present cosmic baryon density parameter $\Omega_{b}$,  and the diffuse baryon fraction $f_\m{d}$~\citep{Connor:2024mjg,Lemos:2025bgy, Li:2020qei, Wang:2022ami, Deng:2013aga, Ravi:2018ose, Munoz:2018mll, Li:2019klc, Walters:2019cie, Wei:2019uhh, Yang:2022ftm, Lin:2023opv, Liu:2025fdf, Zhang:2025yhi, Sales:2026smc, Macquart:2020lln, McQuinn:2013tmc}.  Other cosmological applications have also been investigated~\citep{Muthusami:2026vax,Ravi:2026sem,Ribeiro:2026ucl,Su:2026ejf,takahashi2025measurement}.

For an extragalactic FRB, the observed DM can be decomposed as 
\begin{eqnarray}
	\m{DM}_\m{obs}(z)
	&=&\m{DM}_\m{ISM} + \m{DM}_\m{halo}+ \nonumber\\
	&&\m{DM}_\m{cos}(z) + \m{DM}_\m{host}(z),
\label{eq:comp}
\end{eqnarray}
where the subscripts ISM, halo, cos and host denotes the contributions from the Milky Way interstellar medium, the ionized baryonic gas in the Milky Way halo, the cosmic component from the  intergalactic medium (IGM) together with intervening halos~\citep{Connor:2024mjg}, and the FRB host galaxy, respectively. Since only $\m{DM}_\m{cos}(z)$ directly carries the cosmological information, conventional FRB cosmology requires  subtracting  the contributions of $\m{DM}_\m{ISM}$, $\m{DM}_\m{halo}$ and  $\m{DM}_\m{host}$ from the observed DM.  The Milky Way ISM  contribution is usually estimated using Galactic electron-density models, such as NE2001, YMW16, and NE2025~\citep{Cordes:2002wz,Yao:2017kcp,Ocker:2026mta}. Howver,  different Galactic electron density models can lead to noticeably different constraints on the Hubble constant $H_0$~\citep{Wang:2025ugc}.  

The Galactic halo contribution,  $\m{DM}_\m{halo}$,   is commonly treated either as a fixed value or as a random variable drawn from an assumed distribution. 
However, current observational estimates remain model dependent. 
 Based on constraints derived from the all-sky H{\footnotesize I} 21 cm emission map, O{\footnotesize VII} absorption, and the DM to the Large Magellanic Cloud, together with hydrostatic halo-gas models, 
 $\mathrm{DM_{halo}}$ has been estimated to  fall within in the range of $50\text{--}80$~pc~$\mathrm{cm}^{-3}$.~\citep{prochaska2019probing}.  By contrast,  FRB 20220319D gives a conservative upper limit  of $28.7$ or $47.3$~pc~$\mathrm{cm}^{-3}$~\citep{DeepSynopticArrayTeam:2023suu}.   Analyses of  FRB data at Galactic latitudes $|b| > 30^o$ yields  upper limits on $\mathrm{DM_{halo}}$, ranging from 52 to 111~pc~$\mathrm{cm}^{-3}$~\citep{Cook:2023grs}.  In addition,   a data-driven reconstruction of the all-sky distribution of the dispersion measure contribution from the Galactic halo  indicates that $\mathrm{DM}_{\rm halo}$ may be anisotropic~\citep{Liu:2026txt}.  As a result,  there is currently   no unique or model-independent estimate of  $\m{DM}_\m{halo}$.  Previous studies have also shown that FRB-based cosmological constraints are sensitive to the assumed Galactic-halo DM distribution~\citep{Xu:2025ddk,Liu:2025fdf}. This suggests that adopting a fixed value or a prescribed distribution for $\mathrm{DM}_{\rm halo}$ may introduce systematic uncertainties or biases into the inferred cosmological parameters.

It is therefore important either to model the Milky Way contribution more accurately or to construct observables that are insensitive to it. In this Letter, we propose a differential DM method that eliminates the Milky Way contribution when extracting cosmological information from localized FRBs. The central idea is to compare FRBs located within the same sky region, where the Milky Way ISM and halo contributions are approximately common. Their difference then removes the Milky Way term. We validate the method using simulated FRB samples and then apply it to current localized FRB data.

\section{method}
\label{sec_method}
To remove the contributions from $\m{DM_{ISM}}$ and $\m{DM_{halo}}$, we assume that both vary only weakly within a sufficiently small sky region.  For two localized FRBs in the same region, 
we take the FRB with the lower redshift as the reference event and subtract its observed DM from that of the higher-redshift event. This gives
\begin{eqnarray}
	\label{delta_dm'}
	\Delta \m{DM_{obs}} &=& \m{DM_{obs,2}} - \m{DM_{obs,1}} \nonumber \\
	&=&  \m{DM_{cos,2}}-\m{DM_{cos,1}} +
	\m{DM_{host,2}} - \m{DM_{host,1}} \nonumber\\ 
	&=&  \Delta \m{DM_{cos}} + \Delta \m{DM_{host}}. 
\end{eqnarray}
The resulting quantity is independent of the Milky Way ISM and halo contributions, provided that these contributions are approximately the same for the two lines of sight.

To apply $\Delta \m{DM_{obs}}$ for cosmological inference, we construct its likelihood function as  
\begin{eqnarray}
	\label{p_all}
	P(\Delta \m{DM_{obs}}) 
	&=& \int_{-\infty}^{\infty} P_1(\Delta \m{DM_{cos}}) \nonumber\\
	&&P_2(\Delta \m{DM_{obs}}- \Delta \m{DM_{cos}}) ~\m{d} \Delta \m{DM_{cos}} .
\end{eqnarray}
Since $\Delta \m{DM_{cos}}= \m{DM_{cos,2}}-\m{DM_{cos,1}}$, the integrand $P_1(\Delta \m{DM_{cos}})$ is
\begin{eqnarray}
	\label{p_delta_cos}
	&&P_1(\Delta \m{DM_{cos}}) = \int_{0}^{\infty} P_\m{cos,2}(\m{DM_{cos,2}})\nonumber\\ &&~~~~~~~~~~~~~~~~~~~~P_\m{cos,1}(\m{DM_{cos,2}}- \Delta \m{DM_{cos}}) ~\m{d}\m{DM_{cos,2}} \nonumber\\
	&&= \frac{1}{\langle\m{DM_{cos,1}}\rangle \langle\m{DM_{cos,2}}\rangle }\int_{0}^{\infty} P_\m{\Delta_2}\left( \Delta_2 \right) 
	P_\m{\Delta_1} \left ( \Delta_1 \right)~\m{d}\m{DM_{cos,2}}, \nonumber\\
\end{eqnarray}
where  $\Delta_i \equiv \m{DM_{cos,i}}/\m{\langle DM_{cos,i}\rangle}$  and 
\begin{eqnarray}
	\label{dmcos_aver}
	\m{\langle{DM_{cos}}\rangle}&=&\frac{3 c \Gamma}{8\pi G m_p}
	\int_0^z\frac{(1+z')\chi_e(z')}{E(z')}\m{d}z' . 
	\label{aver_dmcos}
\end{eqnarray}
Here $\langle \cdots \rangle$  represents the average value,  $c$ is the speed of light in vacuum, $G$ is the gravitational constant, $m_p$ is the proton mass, and    $\Gamma \equiv \Omega_{b}  {H_0}f_\m{{d}}$ combines the present-day baryon density fraction $\Omega_b$, the Hubble constant $H_0$, and the diffuse baryon fraction $f_{\rm d}$,  which are degenerate in the mean cosmic DM.  The quantity  $\chi_{e}$ represents the number of free electrons per baryon. Since hydrogen and helium are fully ionized  in the redshift range $z<3$, as indicated by Ly$\alpha$ forest observations~\citep{Becker:2010cu, Meiksin:2007rz},  we take $\chi_{e}=7/8$. In Eq.~(\ref{dmcos_aver}),  $E(z)$ is the dimensionless Hubble parameter. 
 For a spatially flat $\Lambda$CDM model,  $E(z)=\sqrt{\Omega_{m} (1+z)^3 + (1-\Omega_{m}) }$ , where $\Omega_{m}$ is the present matter  density fraction. 

For the distribution of $\Delta_i$, we adopt the quasi-Gaussian probability distribution function 
\begin{eqnarray}
\label{pdelta}	P_{\m{\Delta_i}}(\Delta_i)=A_i\Delta_i^{-\beta_i}\exp\left[-\dfrac{(\Delta_i^{-\alpha_i}-1)^2}{2\alpha_i^2\sigma_{\Delta_i}^2}\right],\Delta_i>0 
\end{eqnarray}
as used in~\citep{McQuinn:2013tmc, prochaska2019probing, Macquart:2020lln,Konietzka:2025kdr, Zhang:2025rvu}.   The parameter $A_i$, $\alpha_i$, $\beta_i$ and $\sigma_{\Delta_i}$ at the redshift $z_i$ are obtained via interpolation of the simulation results reported in~\citep{Zhang:2025rvu}. We note that Ref.~\citep{Macquart:2020lln}  adopted $\alpha=\beta=3$. 

Similarly, the distribution of the host-galaxy DM difference is
\begin{eqnarray}
	\label{p_delta_host}
	P_2(\Delta \m{DM_{host}}) 
	&=&\int_{0}^{\infty} P_\m{host,2}(\m{DM_{host,2}})\cdot \nonumber\\
 &&P_\m{host,1}(\m{DM_{host,2}} - \Delta \m{DM_{host}}) ~\m{d} \m{DM_{host,2}},\nonumber\\
\end{eqnarray}
Following~\citep{Macquart:2020lln, Zhang:2020mgq}, we assume that  $P_\m{host,i}$ satisfies  a log-normal distribution function 
\begin{eqnarray}
\label{phost}
	P_\m{host,i}(\m{DM_{host,i}})&&=\dfrac{1}{\sqrt{2\pi}\m{DM_{host,i}\sigma_{host,i}}}\nonumber\\
	&&\times \exp\m{\left[-\dfrac{(\ln{DM_{host,i}}-\mu_{host,i})^2}{2\sigma_{host,i}^2}\right]},
\end{eqnarray}
where $\m{\mu_{host,i}}$ and $\m{\sigma_{host,i}}$ characterize the host-galaxy DM distribution.  Their values  at the redshift of each FRB are obtained by interpolating the redshift-dependent results from Zhang et al.~\citep{Zhang:2020mgq} based on the IllustrisTNG simulation.

Substituting Eqs.~(\ref{phost}, \ref{p_delta_cos}) into Eq.~(\ref{p_all}) gives the likelihood function of $P(\Delta \m{DM_{obs}})$. The log-likelihood,
$\ln \mathcal{L} = \ln P(\Delta \m{DM_{obs}})$, can then be used to constrain cosmological parameters such as $\Gamma$ through Markov Chain Monte Carlo analysis. Before applying the method to observed FRBs, we first test its reliability using mock data.


\section{simulation}
\label{sec_simu}
Our method is based on the assumption that   both $\m{DM_{ISM}}$ and $\m{DM_{halo}}$  vary  negligibly  within each adopted sky region. We first test the validity of this assumption.

We generate 64,800 mock sky positions uniformly distributed over the sky  and assign  each position a Milky Way contribution of $\m{DM_{MW}}=\m{DM_{ISM}}+\m{DM_{halo}}$.  The ISM contribution  $\m{DM_{ISM}}$  is generated  using the NE2025  Galactic electron-density model~\citep{Ocker:2026mta},  while the halo contribution  $\m{DM_{halo}}$  is generated using the YT2020 model~\citep{yamasaki2020galactic}.  We exclude the low-Galactic-latitude region $|b|<15^\circ$, where the dense and complex Galactic environment produces large and rapidly varying Milky Way DMs.  We further exclude an overdense region  around $l\approx260^\circ$ and $b\approx-15^\circ$, associated with the Gum and Vela clumps considered in~\citep{Ocker:2026mta}. In this region.  $\m{DM_{MW}}$ exhibits significant spatial variation. None of observed localized FRBs used below lies within this excluded region.

\begin{figure*}[htbp]
	\centering
	\includegraphics[width=1\textwidth]{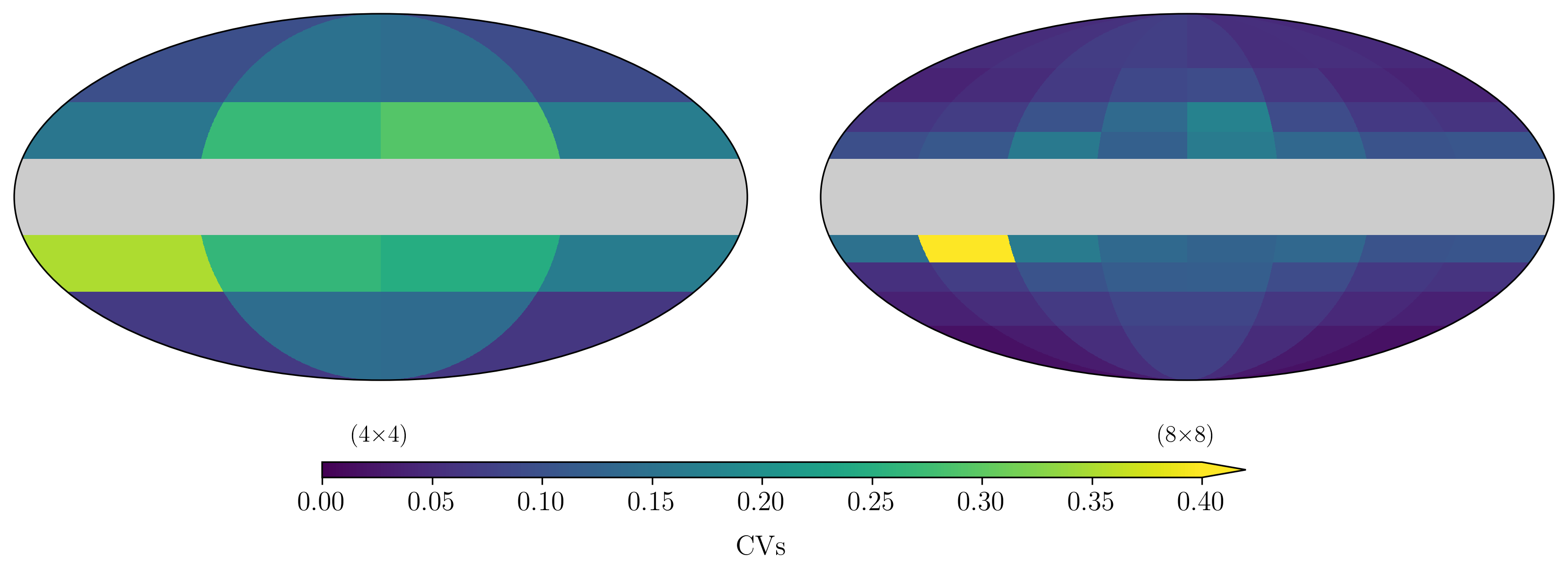}
	\caption{Coefficients of variation (CVs) of $\m{DM_{MW}}$ obtained under different sky-partitioning schemes. The left and right panels show the $4\times4$ and $8\times8$ partitions, respectively. Colors indicate the CV values according to the color bar at the bottom, with all values exceeding 0.4 shown in yellow. The gray band marks the excluded Galactic plane region, corresponding to $|b|<15^\circ$. The overdense region in the 8$\times$8 partition is associated with the Gum and Vela clumps, whose CV values exceed 0.4.}
		\label{fig_cv}
\end{figure*}

To quantify the spatial variation of $\m{DM_{MW}}$, we employ the coefficient of variation (CV), defined as $\m{CV} \equiv \sigma(\m{DM_{MW}}) / \mu(\m{DM_{MW}})$,  where $\sigma$ and $\mu$ denote the standard deviation and mean of $\m{DM_{MW}}$ within a given sky region, respectively. A smaller CV indicates a more homogeneous distribution of $\m{DM_{MW}}$. We adopt ${\rm CV}<0.2$ as the criterion for approximate homogeneity.  We then examine different sky-partitioning schemes.  Starting with the full sky,   we divide the Galactic longitude range $0^\circ$-$360^\circ$ and Galactic latitude range $-90^\circ$-$90^\circ$ into $2\times2$, $4\times4$, and $8\times8$ bins, corresponding to 4, 16, and 64 sky regions, respectively.  
Without partitioning, and also in the $2\times2$ scheme,  the CV values greatly exceed $0.2$, indicating substantial variations in $\m{DM_{MW}}$. In the  $4\times4$ svhem, most regions still have $\m{CV}>0.2$. In the $8\times8$ scheme, all regions satisfy the criterion $\m{CV}<0.2$, except for the overdense Gum-Vela region discussed above.   These results motivate the use of an $8\times8$ sky partition. The CV distributions for the $4\times4$ and $8\times8$ schemes are shown in Fig.~\ref{fig_cv}.
  
We next test the performance of the method using mock FRB samples. We generate 200 mock FRBs, comparable to the current number of localized FRBs~\citep{Sales:2026smc}.  The mock FRBs are distributed over the sky outside the region $|b|<15^\circ$, and their redshift distribution follows~\citep{Qiang:2021bwb}
\begin{eqnarray}
	P_\m{model}(z) \propto z \exp(-z),
\end{eqnarray}
with an upper  redshift cutoff  at $z = 1.5$.  For each mock FRB, the observed DM,   $\m{DM_{obs}}$, is constructed as
\begin{eqnarray}
	\m{DM_{obs,sim}} &=& \m{DM_{ISM,sim}} + \m{DM_{halo,sim}} \nonumber\\
	&&+ \m{DM_{cos,sim}} + \m{DM_{host,sim}}.
\end{eqnarray}
Here, $\m{DM_{ISM,sim}}$ and $\m{DM_{halo,sim}}$ are generated using the NE2025  model and the YT2020 model, respectively.
The $\m{DM_{cos,sim}}$ is generated from $\m{DM_{cos,sim}} = \langle \m{DM_{cos,sim}}\rangle \cdot \Delta_\m{sim}$,
where $\Delta_\m{sim}$ is drawn from the PDF  in Eq.~(\ref{pdelta}), and $\langle \m{DM_{cos,sim}}\rangle$ is calculated from Eq.~(\ref{dmcos_aver}) within the framework of the $\Lambda$CDM model. Here we adopt  $\Omega_{m}=0.315$ and  $\Gamma = 3.12$, where the latter is obtained from  $\Omega_{b} h^2 = 0.02237$, $H_0 = 100h~\m{km~s^{-1}~Mpc^{-1}}=67.36~\m{km~s^{-1}~Mpc^{-1}}$, and $f_\m{d} = 0.94$. The  values of    $\Omega_{m}$, $\Omega_{b} h^2$ and  $H_0$ are taken from Planck 2018 CMB results~\citep{Planck:2018vyg}, while $f_\m{d}$ is adopted from~\citep{Connor:2024mjg}.
The simulated host-galaxy contribution, $\m{DM_{host,sim}}$, is generated from Eq.~(\ref{phost}), with  $\mu_\mathrm{host}$ and $\sigma_\mathrm{host}$ obtained via interpolation at the corresponding redshift.

Now Applying our differential method to the simulated $\mathrm{DM}_{\rm obs}$ data, we infer the posterior distribution of $\Gamma$. To reduce statistical fluctuations from any single simulation, we  repeat simulation and inference procedure 100 times, yielding $\Gamma = 3.10\pm 0.19$.  This result  is  consistent with the fiducial value of $\Gamma=3.12$ used in the simulation. This agreement demonstrates that the method can reliably recover the input cosmological parameter under the assumptions adopted in the simulation.
 
\section{constraint  on $\Gamma$ from Localized FRBs}
\label{sec_real}

\begin{figure}[H]
	\centering
	\includegraphics[width=0.45\textwidth]{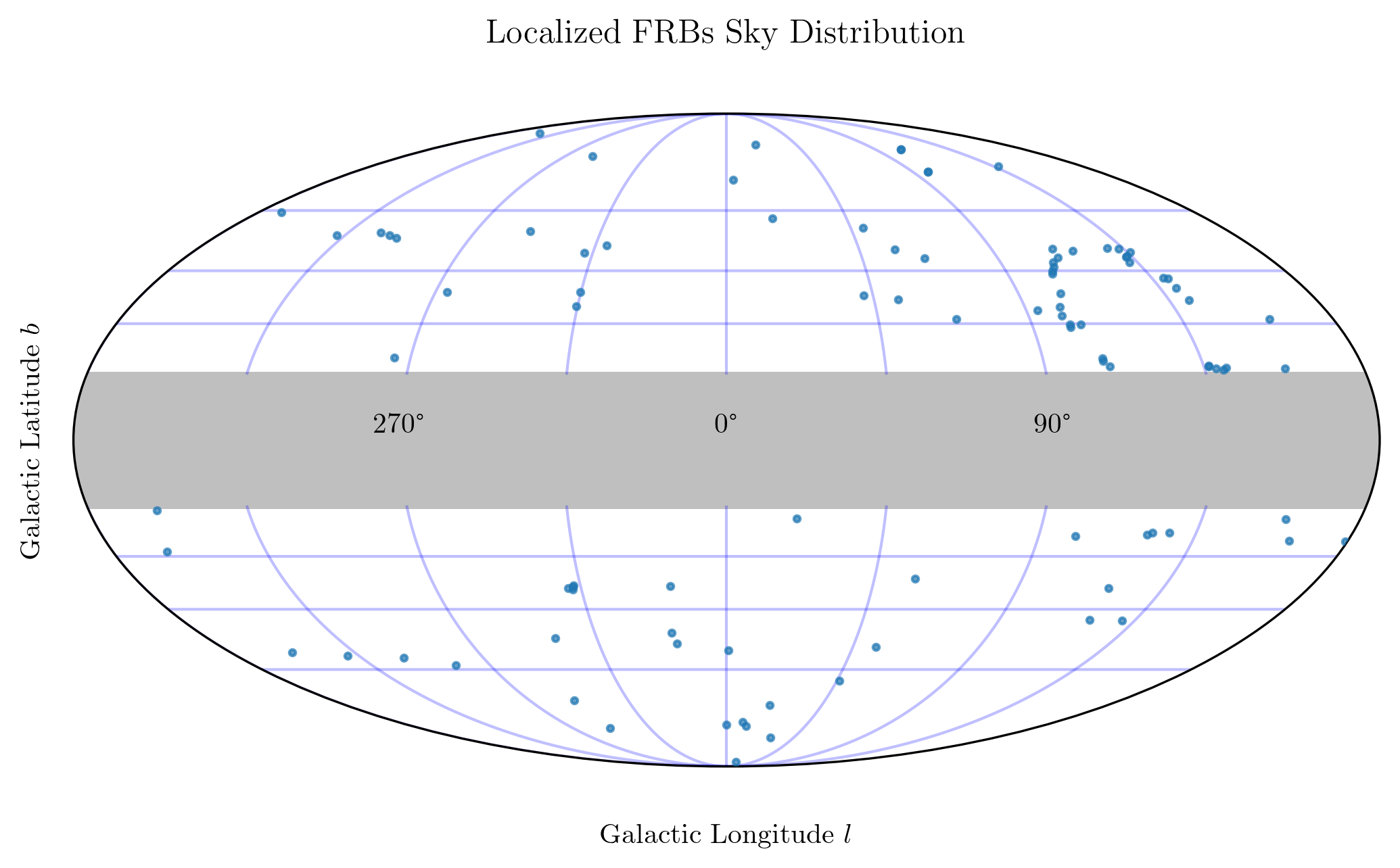}
	\caption{Sky distribution of the 99 localized FRB samples. The blue points represent the FRB sample locations, while the light blue lines partition the sky into 64 regions. The gray shaded band indicates the excluded Galactic plane region with $|b| < 15^\circ$.
	}
	\label{fig_loc}
\end{figure}

We now  apply the method to  127 observed localized FRBs,  listed in Appendix.~\ref{frbdata}. In our analysis,   we exclude FRB20190520B and FRB20220831A  because their unusually large DMs are likely dominated by the local source environments~\citep{Wu:2021jyk,Connor:2024mjg}. We also exclude  FRB20221027A because of ambiguity in its host-galaxy association~\citep{Sharma:2024fsq}.  Furthermore, we remove FRBs with redshift $z < 0.02$ or Galactic latitude $|b| < 15^\circ$~\citep{Zhang:2025rvu,Liu:2026txt}. Finally,   one FRB with redshift $z > 1.5$ is excluded, because the simulations used to calibrate the host-galaxy DM distribution extend only to $z=1.5$~\citep{Zhang:2020mgq}. This leaves 99 FRBs  in the final sample.  Their sky distribution is shown in Fig.~\ref{fig_loc}.  

We use the $8\times8$ sky-partitioning scheme established above. In each sky region, the lowest-redshift FRB is taken as the reference event, corresponding to $\m{DM_{obs,1}}$,  while the remaining FRBs in the same region are used as $\m{DM_{obs,2}}$.  We then construct $\Delta \mathrm{DM}_{\rm obs}$ and use the likelihood described in Sec.~\ref{sec_method} to constrain $\Gamma$.

\begin{figure}[H]
	\centering
	\includegraphics[width=0.45 \textwidth]{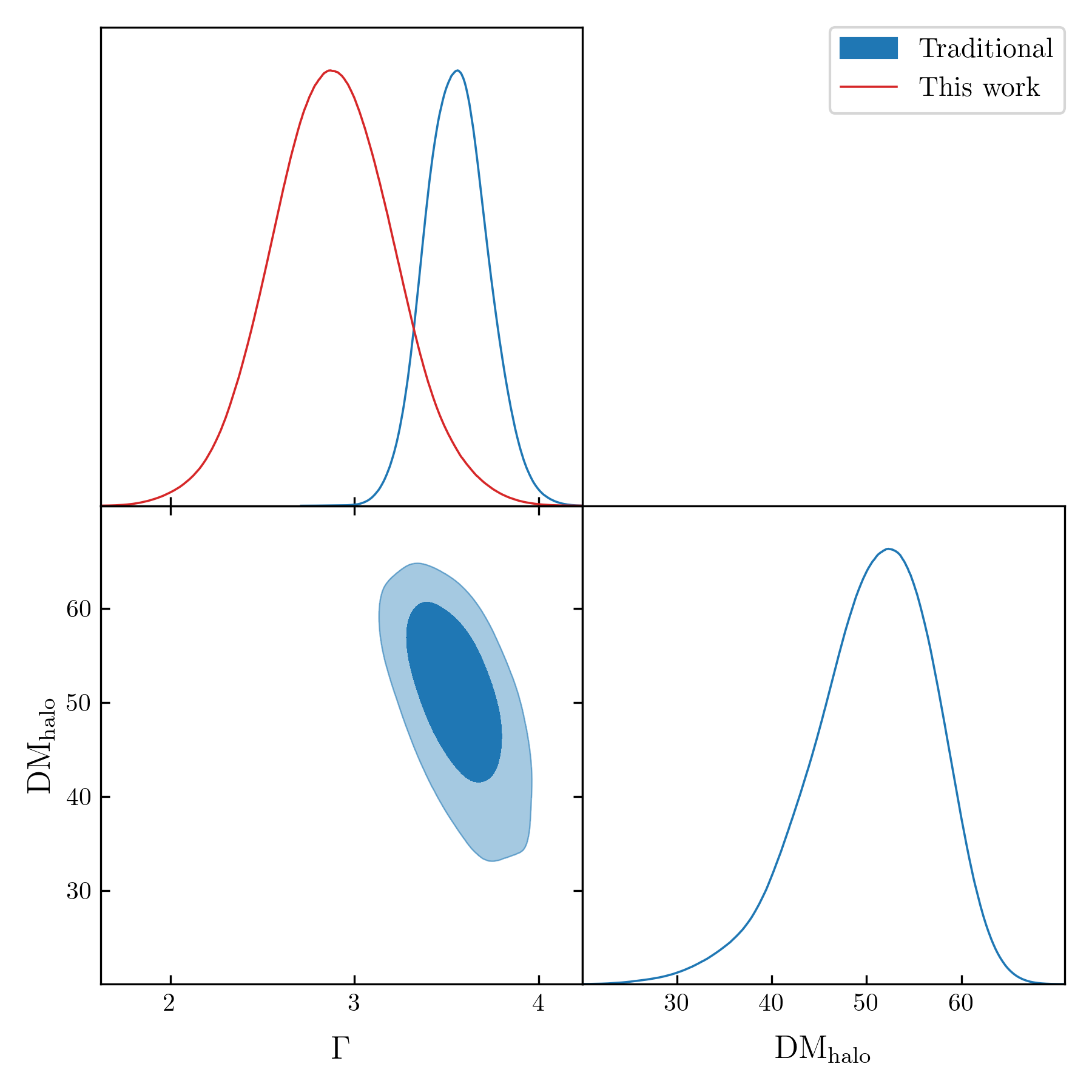}
	\caption{Posterior distribution of $\Gamma$ obtained with the method proposed in this paper (red) and with the traditional method (blue). The proposed method yields $\Gamma=2.88\pm0.33$, while the traditional method gives $\Gamma=3.55\pm0.17$ and $\m{DM_{halo}}=51^{+8}_{-5}~\m{pc~cm^{-3}}$.}
	\label{fig_real}
\end{figure}

After performing the MCMC analysis using the \texttt{emcee} package~\citep{Foreman-Mackey:2012any}, we obtain $\Gamma=2.88\pm0.33$. The corresponding posterior distribution is presented by the red curve in Fig.~\ref{fig_real}. Fixing $\Omega_b h^2 = 0.02237$ and $H_0=67.36~\m{km~s^{-1}~Mpc^{-1}}$,   we find $f_\m{d}=0.867\pm0.099$  from the constraint on   $\Gamma$. 
For comparison, we also analyze the same FRB sample using the conventional method. In this case,  the ISM contribution is subtracted using the NE2025 model, while $\m{DM_{halo}}$ is treated as a free parameter.  The resulting posterior distribution is shown by the blue curve in Fig.~\ref{fig_real}.  We find $\Gamma=3.55\pm0.17$ and $\m{DM_{halo}}=51^{+8}_{-5}~\m{pc~cm^{-3}}$.  With the same fixed values of $\Omega_b h^2$ and $H_0$,  this corresponds to $f_\m{d}=1.069\pm0.051$.  The value of $\Gamma$ inferred from our differential method is lower than that obtained using the conventional method by about $1.8\sigma$. 
This difference suggests that the treatment of $\mathrm{DM_{MW}}$ can have a significant impact on cosmological parameter estimation from FRBs.  This conclusion is consistent with previous studies showing that different Galactic electron-density models and different assumptions about the Galactic halo contribution can lead to noticeably different constraints on $H_0$ or $f_{\rm d}$~\citep{Wang:2025ugc,Xu:2025ddk,Liu:2025fdf}.


\section{Conclusions and discussions}
\label{sec_conclusion}

Localized FRBs provide a promising probe of cosmology through the relation between DM and redshift. 
However, extracting the cosmological DM component from the observed DM requires accounting for several foreground and environmental contributions. In particular, the Milky Way ISM and halo contributions are usually estimated using Galactic electron-density models and assumed halo-DM prescriptions. The uncertainties in these terms propagate directly into the inferred cosmic DM and can therefore bias cosmological constraints.

In this Letter, we have proposed a differential method that removes the Milky Way contribution without relying on a specific Galactic electron-density model or a prior assumption about $\m{DM_{halo}}$. The method compares localized FRBs within the same sky region and uses their DM differences. Under the assumption that the Milky Way contribution varies weakly within each region, the ISM and halo terms cancel in the difference, leaving a quantity sensitive only to the cosmic and host-galaxy DM differences.

Using mock FRB samples, we tested the reliability of the method. For simulations generated with the fiducial value $\Gamma=3.12$, the method recovers $\Gamma=3.10\pm0.19$, demonstrating that it can accurately infer the input cosmological parameter under controlled conditions. Applying the method to current localized FRB data, we obtain $\Gamma=2.88\pm0.33$, corresponding to $f_{\m d}=0.867\pm0.099$ when $\Omega_b h^2$ and $H_0$ are fixed to their Planck 2018 values. By contrast, the conventional method applied to the same sample gives $\Gamma=3.55\pm0.17$ and $\mathrm{DM}_{\rm halo}=51^{+8}_{-5}~\m{ pc~cm^{-3}}$, corresponding to $f_{\m d}=1.069\pm0.051$. The discrepancy between the two approaches indicates that the treatment of the Milky Way DM contribution is a non-negligible systematic in current FRB cosmology.

The method proposed here provides a way to reduce systematic uncertainties associated with the Milky Way foreground. Its statistical precision is currently limited by the number and sky distribution of localized FRBs, since each sky region requires at least one low-redshift reference event and additional higher-redshift events. As the localized FRB sample grows, the differential approach developed here will become increasingly powerful and may offer a robust route to precision cosmology with FRBs.

\begin{acknowledgments}
We appreciate  Prof. Xuebing Wu for the enlightening discussions. This work was supported by the National Natural Science Foundation of China (Grants  No.~12275080 and No.~12075084), the Major basic research project of Hunan Province (Grant No.~2024JC0001), and the Innovative Research Group of Hunan Province (Grant No.~2024JJ1006).
\end{acknowledgments}

\bibliography{ref}{}
\bibliographystyle{aasjournalv7}

\appendix
\section{FRB data}
\label{frbdata}

\begin{table}[H]
	\centering
\begin{tabular}{ccccccc}
	\toprule
	FRB Type & TNS Name & l & b & $\m{DM_{obs}}(\m{pc~cm^{-3}})$ ~~& Redshift ~~&~~ Reference\\
	\midrule
	1 & FRB20121102A & 174.8895 & -0.2313 & 557 & 0.1927 &\cite{Tendulkar:2017vuq}\\
	&&&&&&\cite{Chatterjee:2017dqg}\\ 
	3 & FRB20171020A & 36.358 & -53.5548 & 114.1 & 0.0087 & \cite{Mahony:2018ddp}\\
	1 & FRB20180301A & 204.3368 & -6.3899 & 536 &  0.3304 & \cite{Bhandari:2021pvj}\\
	2 & FRB20180814A & 136.4708 & 16.6458 & 189.4 & 0.068 & \cite{Michilli:2022bbs}\\
	2 & FRB20180916B & 129.7108 & 3.725 & 348.76 &  0.0337 & \cite{Marcote:2020ljw}\\
	3 & FRB20180924B & 0.7424 & -49.4148 & 361.42 &  0.3214 & \cite{Bannister:2019iju}\\
	1 & FRB20181030A & 134.8128 & 40.0569 & 103.5 &  0.0039 & \cite{Bhardwaj:2021hgc} \\
	3 & FRB20181112A & 342.5995 & -47.6988 & 589.27 &  0.4755 & \cite{prochaska2019low}\\
	3 & FRB20181220A & 106.825 & -11.4882 & 209.4 & 0.0275 & \cite{Bhardwaj:2023vha}\\
	3 & FRB20181223C & 207.9002 & 79.398 & 112.5 &  0.0302 & \cite{Bhardwaj:2023vha}\\
	3 & FRB20190102C & 312.6537 & -33.4931 & 364.5 &  0.2913 & \cite{Bhandari:2020oyb}\\
	2 & FRB20190110C & 65.5731 & 42.1 & 221.6 &  0.1224 & \cite{Ibik:2023ugl}\\
	1 & FRB20190303A & 97.6424 & 65.9258 & 222.4 &  0.064 & \cite{Michilli:2022bbs}\\
	3 & FRB20190418A & 179.2967 & -22.892 & 184.5 &  0.0713 & \cite{Bhardwaj:2023vha}\\
	3 & FRB20190425A & 42.1039 & 33.0787 & 128.2 &  0.0312 & \cite{Bhardwaj:2023vha}\\
	1 & FRB20190520B & 359.6726 & 29.9074 & 1210.3 &  0.241 & \cite{Niu:2021bnl} \\
	3 & FRB20190523A & 117.0338 & 44.0019 & 760.8 &  0.66 & \cite{Ravi:2019alc} \\
	3 & FRB20190608B & 53.2088 & -48.5295 & 339.5 &  0.1178 & \cite{chittidi2021dissecting}\\
	3 & FRB20190611B & 312.9352 & -33.2818 & 321.4 &  0.378& \cite{heintz2020host}  \\
	3 & FRB20190614D & 136.3059 & 16.54 & 959.2 &  0.6 & \cite{Law:2020cnm}\\
	1 & FRB20190711A & 310.9081 & -33.902 & 593.1 &  0.522 & \cite{heintz2020host} \\
	3 & FRB20190714A & 289.6971 & 48.9358 & 504.13 &  0.2365 & \cite{heintz2020host} \\
	3 & FRB20191001A & 341.2276 & -44.904 & 507.9 &  0.234 & \cite{heintz2020host}\\
	1 & FRB20191106C & 105.6894 & 73.217 & 332.2  & 0.1077 & \cite{Ibik:2023ugl}\\
	3 & FRB20191228A & 20.5554 & -64.9244 & 297.5  & 0.2432 & \cite{Bhandari:2021pvj}\\
	3 & FRB20200120E & 142.1975 & 41.2167 & 87.8  & 0.0008 & \cite{Kirsten:2021llv} \\
	2 & FRB20200223B & 118.0886 & -33.8678 & 201.8  & 0.0602 & \cite{Ibik:2023ugl}\\
	3 & FRB20200430A & 17.1404 & 52.5033 & 380.25  & 0.16 & \cite{heintz2020host}\\
	3 & FRB20200906A & 202.257 & -49.9984 & 577.8 & 0.3688 & \cite{Bhandari:2021pvj} \\
	3 & FRB20201123A & 340.3141 & -9.6311 & 433.55  & 0.0507 & \cite{Rajwade:2022zkj}\\
	2 & FRB20201124A & 177.7677 & -8.5231 & 413.52 &  0.098 & \cite{Ravi:2021kqk} \\
	3 & FRB20210117A & 45.9174 & -57.6465 & 728.95 &  0.214 & \cite{Bhandari:2022ton} \\
	3 & FRB20210320C & 318.8729 & 45.3081 & 384.8 &  0.2797 & \cite{Shannon:2024pbu} \\
	3 & FRB20210405I & 338.192 & -4.5968 & 565.17 &  0.066 & \cite{Driessen:2023lxj} \\
	3 & FRB20210410D & 312.3222 & -34.1294 & 578.78 &  0.1415 & \cite{Caleb:2023atr} \\
	3 & FRB20210603A & 119.7095 & -41.5808 & 500.147 &  0.177 & \cite{Cassanelli:2023hvg} \\
	3 & FRB20210807D & 39.8612 & -14.8778 & 251.3 & 0.1293 & \cite{Gordon:2023cgw}\\
	&&&&&& \cite{Shannon:2024pbu} \\

	\bottomrule
\end{tabular}
\end{table}

\begin{table}
	\centering
\begin{tabular}{ccccccc}
	\toprule
	FRB Type & TNS Name & l & b & $\m{DM_{obs}}(\m{pc~cm^{-3}})$ ~~& Redshift $z$ & Reference\\
	\midrule
	3 & FRB20211127I & 312.0214 & 43.5427 & 234.83 &  0.0469 & \cite{Gordon:2023cgw}\\
	&&&&&& \cite{Shannon:2024pbu}\\
	&&&&&& \cite{glowacki2023wallaby} \\
	3 & FRB20211203C & 314.5185 & 30.4363 & 635 &  0.3439 & \cite{Gordon:2023cgw} \\
	3 & FRB20211212A & 244.0081 & 47.3154 & 206 & 0.0715 & \cite{Gordon:2023cgw}\\
	&&&&&& \cite{Shannon:2024pbu} \\
	3 & FRB20220105A & 18.5567 & 74.8076 & 580 & 0.2785 & \cite{Gordon:2023cgw} \\
	3 & FRB20220204A & 100.0311 & 28.2958 & 612.2 &  0.4 & \cite{Sharma:2024fsq} \\
	3 & FRB20220207C & 106.9385 & 18.3896 & 262.38 &  0.043 & \cite{Law:2023ibd} \\
	3 & FRB20220208A & 107.2615 & 13.5609 & 437 &  0.351 & \cite{Sharma:2024fsq} \\
	3 & FRB20220307B & 116.2459 & 10.4718 & 499.27 &  0.2481 & \cite{Law:2023ibd} \\
	3 & FRB20220310F & 140.0244 & 34.798 & 462.24 &  0.478 & \cite{Law:2023ibd} \\
	3 & FRB20220319D & 129.1829 & 9.1073 & 110.95 &  0.0111 & \cite{DeepSynopticArrayTeam:2023suu} \\
	3 & FRB20220330D & 136.0092 & 43.7016 & 468.1 &  0.3714 & \cite{Sharma:2024fsq} \\
	3 & FRB20220418A & 110.7531 & 44.4672 & 623.25 &  0.622 & \cite{Law:2023ibd} \\
	3 & FRB20220501C & 11.1777 & -71.4731 & 449.5 &  0.381 & \cite{Shannon:2024pbu} \\
	3 & FRB20220506D & 108.3544 & 16.5134 & 396.97 &  0.3004 & \cite{Law:2023ibd} \\
	3 & FRB20220509G & 100.9432 & 25.4764 & 269.53 &  0.0894 & \cite{Law:2023ibd} \\
	1 & FRB20220529A & 130.7876 & -41.8579 & 246 &  0.1839 & \cite{Gao:2025fcr} \\
	3 & FRB20220610A & 8.8396 & -70.1857 & 1457.624 &  1.016 & \cite{Ryder:2022qpg} \\
	3 & FRB20220717A & 19.8352 & -17.6321 & 637 &  0.3629 & \cite{Rajwade:2024ozu} \\
	3 & FRB20220725A & 0.0017 & -71.1863 & 290.4 & 0.1926 & \cite{Shannon:2024pbu} \\
	3 & FRB20220726A & 141.1708 & 16.2971 & 686.55 &  0.361 & \cite{Sharma:2024fsq} \\
	3 & FRB20220825A & 106.995 & 17.7852 & 651.24 &  0.2414 & \cite{Law:2023ibd} \\
	3 & FRB20220831A & 110.96 & 12.47 & 1146.25 &  0.262 & \cite{Connor:2024mjg} \\
	2 & FRB20220912A & 106.0651 & -10.7838 & 219.46 &  0.0771 & \cite{DeepSynopticArrayTeam:2022rbq} \\
	3 & FRB20220914A & 104.308 & 26.1301 & 631.28 & 0.1139 & \cite{Law:2023ibd} \\
	3 & FRB20220918A & 300.6851 & -46.2342 & 656.8 & 0.491 & \cite{Shannon:2024pbu} \\
	3 & FRB20220920A & 104.9231 & 38.8933 & 314.99 &  0.1582 & \cite{Law:2023ibd} \\
	3 & FRB20221012A & 101.1451 & 26.1399 & 441.08 & 0.2847 & \cite{Law:2023ibd} \\
	3 & FRB20221027A & 142.088 & 34.2206 & 452.5 &  0.229 & \cite{Sharma:2024fsq} \\
	3 & FRB20221029A & 140.0034 & 37.1576 & 1391.05 & 0.975  & \cite{Sharma:2024fsq}\\
	3 & FRB20221101B & 113.0672 & 10.1947 & 490.7 &  0.2395 & \cite{Sharma:2024fsq} \\
	3 & FRB20221106A & 220.901 & -50.8788 & 343.8 &  0.2044 & \cite{Shannon:2024pbu} \\
	3 & FRB20221113A & 140.3183 & 15.8337 & 411.4 &  0.2505 & \cite{Sharma:2024fsq} \\
	3 & FRB20221116A & 125.4515 & 9.9458 & 640.6 &  0.2764 & \cite{Sharma:2024fsq} \\
	3 & FRB20221219A & 102.9308 & 33.5259 & 706.7 &  0.554 & \cite{Sharma:2024fsq} \\
	3 & FRB20230124A & 107.127 & 41.0275 & 590.6 & 0.094 & \cite{Sharma:2024fsq} \\
	3 & FRB20230203A & 188.7125 & 54.087 & 420.1 &  0.1464  & \cite{CHIMEFRB:2025ggb}\\
	3 & FRB20230216A & 240.7961 & 47.9456 & 828 &  0.531 & \cite{Sharma:2024fsq} \\
	3 & FRB20230222A & 204.7164 & 8.6957 & 706.1 &  0.1223 & \cite{CHIMEFRB:2025ggb} \\
	3 & FRB20230222B & 49.6176 & 49.9787 & 187.8 &  0.11 & \cite{CHIMEFRB:2025ggb} \\
	3 & FRB20230307A & 129.5267 & 44.6523 & 608.9 &  0.271 & \cite{Sharma:2024fsq} \\
	\bottomrule
\end{tabular}
\end{table}

\begin{table}
		\centering
\begin{tabular}{ccccccc}
	\toprule
	FRB Type & TNS Name & l & b & $\m{DM_{obs}}(\m{pc~cm^{-3}})$ ~~& Redshift $z$ & Reference\\
	\midrule		
	3 & FRB20230311A & 157.7134 & 16.0395 & 364.3 &  0.1918 & \cite{CHIMEFRB:2025ggb} \\
	3 & FRB20230501A & 112.5401 & 10.7493 & 532.5 &  0.301 & \cite{Sharma:2024fsq} \\
	1 & FRB20230506C & 122.3277 & -20.8621 & 766.5 &  0.3896 & \cite{Anna-Thomas:2025zao} \\
	3 & FRB20230521B & 115.9336 & 9.462 & 1342.9 & 1.354 & \cite{Connor:2024mjg} \\
	3 & FRB20230526A & 290.171 & -63.4721 & 361.4 & 0.157 & \cite{Shannon:2024pbu} \\
	3 & FRB20230626A & 106.2755 & 39.997 & 451.2 &  0.327 & \cite{Sharma:2024fsq} \\
	3 & FRB20230628A & 133.4117 & 42.6599 & 345.15 &  0.1265 & \cite{Sharma:2024fsq} \\
	3 & FRB20230703A & 137.2098 & 67.475 & 291.3 & 0.1184 & \cite{CHIMEFRB:2025ggb} \\
	3 & FRB20230708A & 342.6288 & -33.3877 & 411.51 &  0.105 & \cite{Shannon:2024pbu} \\
	3 & FRB20230712A & 133.0151 & 42.5161 & 586.96 &  0.4525 & \cite{Sharma:2024fsq} \\
	3 & FRB20230718A & 259.4629 & -0.3666 & 477 & 0.035 & \cite{Shannon:2024pbu} \\
	3 & FRB20230730A & 158.8166 & -17.8164 & 312.5 &  0.2115 & \cite{CHIMEFRB:2025ggb} \\
	3 & FRB20230814A & 112.5635 & 13.1971 & 696.4 &  0.5535 & \cite{Connor:2024mjg} \\
	3 & FRB20230902A & 256.9906 & -53.3387 & 440.1 &  0.3619 & \cite{Shannon:2024pbu} \\
	3 & FRB20230926A & 68.2353 & 27.484 & 222.8 &  0.0553 & \cite{CHIMEFRB:2025ggb} \\
	3 & FRB20230930A & 121.0371 & -21.4196 & 456 &  0.0925 & \cite{Anna-Thomas:2025zao} \\
	3 & FRB20231005A & 57.1653 & 44.3532 & 189.4 &  0.0713 & \cite{CHIMEFRB:2025ggb} \\
	3 & FRB20231011A & 127.2287 & -20.943 & 186.3 &  0.0783 & \cite{CHIMEFRB:2025ggb} \\
	3 & FRB20231017A & 100.6098 & -21.6519 & 344.2 &  0.245 & \cite{CHIMEFRB:2025ggb} \\
	3 & FRB20231025B & 93.4332 & 29.434 & 368.7 &  0.3238 & \cite{CHIMEFRB:2025ggb} \\
	3 & FRB20231120A & 138.7226 & 37.2339 & 438.9 &  0.07 & \cite{Sharma:2024fsq} \\
	3 & FRB20231123A & 199.3094 & -15.7427 & 302.1 &  0.0729 & \cite{CHIMEFRB:2025ggb} \\
	3 & FRB20231123B & 104.2782 & 38.3244 & 396.7 &  0.2625 & \cite{Sharma:2024fsq} \\
	3 & FRB20231201A & 163.0405 & -22.7702 & 169.4 &  0.1119 & \cite{CHIMEFRB:2025ggb} \\
	3 & FRB20231206A & 161.0582 & 27.4819 & 457.7 &  0.0659 & \cite{CHIMEFRB:2025ggb} \\
	3 & FRB20231220A & 140.9346 & 31.8365 & 491.2 &  0.3355 & \cite{Connor:2024mjg} \\
	3 & FRB20231223C & 52.3107 & 32.0802 & 165.8 &  0.1059 & \cite{CHIMEFRB:2025ggb} \\
	3 & FRB20231226A & 236.5768 & 48.6458 & 329.9 &  0.1569 & \cite{Shannon:2024pbu} \\
	3 & FRB20231229A & 135.3449 & -26.4433 & 198.5 &  0.019 & \cite{CHIMEFRB:2025ggb} \\
	1 & FRB20240114A & 57.481 & -31.6512 & 527.7 &  0.13 & \cite{Tian:2024ygd} \\
	3 & FRB20240119A & 110.0349 & 42.2065 & 483.1 &  0.37 & \cite{Connor:2024mjg} \\
	3 & FRB20240123A & 138.3287 & 16.0762 & 1462 &  0.968 & \cite{Connor:2024mjg} \\
	3 & FRB20240201A & 222.1335 & 47.9692 & 374.5 &  0.0427 & \cite{Shannon:2024pbu} \\
	3 & FRB20240210A & 14.4396 & -86.2116 & 283.73 &  0.0237 & \cite{Shannon:2024pbu} \\
	3 & FRB20240213A & 132.3368 & 41.0451 & 357.4 & 0.1185 & \cite{Connor:2024mjg} \\
	3 & FRB20240215A & 100.6441 & 30.2691 & 549.5 &  0.21 & \cite{Connor:2024mjg} \\
	3 & FRB20240229A & 133.3327 & 44.5583 & 491.15 &  0.287 & \cite{Connor:2024mjg} \\
	3 & FRB20240310A & 291.7066 & -72.2713 & 601.8 &  0.127 & \cite{Shannon:2024pbu} \\
	3 & FRB20220224C & 273.6202 & 33.9026 & 1140.2 &  0.6271 & \cite{Pastor-Marazuela:2025loc} \\
	3 & FRB20230907D & 285.5486 & 70.7541 & 1030.79 & 0.4638 & \cite{Pastor-Marazuela:2025loc} \\
	\bottomrule
\end{tabular}
\end{table}

\begin{table}
		\centering
\begin{tabular}{ccccccc}
	\toprule
	FRB Type & TNS Name & l & b & $\m{DM_{obs}}(\m{pc~cm^{-3}})$ ~~& Redshift $z$ & Reference\\
	\midrule
	
	3 & FRB20231020B & 240.4042 & -51.4227 & 952.2 &  0.4775 & \cite{Pastor-Marazuela:2025loc} \\
	3 & FRB20240304B & 269.8675 & 72.1035 & 2458.2 &  2.148 & \cite{Caleb:2025uzd} \\
	3 & FRB20241228A & 3.1431 & 63.2658 & 246.3 & 0.1614 & \cite{Curtin:2025tvg} \\
	3 & FRB20231128A & 105.6852 & 73.224 & 331.6 &  0.1079 & \cite{CHIMEFRB:2025ggb} \\
	3 & FRB20231204A & 97.6238 & 65.9282 & 221 &  0.0644 & \cite{CHIMEFRB:2025ggb} \\
	3 & FRB20231230A & 195.8701 & -25.2387 & 131.4 & 0.0298 & \cite{CHIMEFRB:2025ggb} \\
	3 & FRB20220222C & 314.6791 & 33.8351 & 1071.2 & 0.853 & \cite{Pastor-Marazuela:2025loc} \\
	3 & FRB20230125D & 265.39 & 18.6188 & 640.08 &  0.3265 & \cite{Pastor-Marazuela:2025loc} \\
	3 & FRB20230613A & 29.5697 & -75.7588 & 483.51 &  0.3923 & \cite{Pastor-Marazuela:2025loc} \\
	3 & FRB20250316A & 132.9535 & 57.4605 & 161.82 &  0.0067 & \cite{CHIME:2025mlf} \\
	\bottomrule
\end{tabular}
\end{table}

\end{document}